\documentclass[%
 amsmath,amssymb,
 aps,
]{revtex4-1}

\usepackage{color}
\usepackage[utf8]{inputenc}
\usepackage{graphicx}
\usepackage{dcolumn}
\usepackage{bm}

\begin{document}

\title{Susceptibility of power grids to input fluctuations}

\author{Stefan Wieland}
\email{stefan.wieland@tuta.io}
\author{S\'ebastien Aumaitre}%
\author{Herv\'e Bercegol}
\affiliation{SPEC, CEA, CNRS, Universit\'e Paris-Saclay, F-91191 Gif-sur-Yvette Cedex, France}

\date{\today}

\begin{abstract}
With the increasing inclusion of regenerative resources in the energy mix, their intermittent character challenges power grid stability. Hence it is essential to determine which input fluctuations power grids are particularly vulnerable to. Focusing on angular stability in transmission grids, we propose a linear-response approach that yields a frequency-resolved measure of a grid's susceptibility to temporal input fluctuations. This approach can be applied to arbitrary transmission grid topologies as well as other settings described by oscillator networks. 
\end{abstract}

\pacs{Valid PACS appear here}
\maketitle


\section{Introduction}
Power grids are highly structured networks that account for geographically distributed populations and energy sources. High-voltage transmission grids are designed to bridge large distances from generators to substations, where power is then delegated to local lower-voltage distribution grids. Transmission grids are predominantly operated with alternating currents, and can be considered oscillator networks where generators (injecting power) and consumers (substations drawing power) are coupled rotating machines described by Kuramoto-type equations \cite{kuramoto_chemical_1984,filatrella_analysis_2008,nishikawa_comparative_2015}. Stable grid operations are then reflected by frequency synchronization where all oscillators rotate with the same angular velocity, which is $50$ Hz in Europe as well as large parts of Africa and Asia, and $60$ Hz predominantly in the Americas. In a reference frame co-rotating with that frequency, this synchronized regime is manifest in steady-state (or "locked") oscillator phases \cite{kuramoto_chemical_1984}. Synchronization is found in many other instances of collective dynamics and has been the subject of extensive research (see \cite{acebron_kuramoto_2005,arenas_synchronization_2008,dorfler_synchronization_2014,gupta_kuramoto_2014,rodrigues_kuramoto_2016}  for reviews).

Perturbations of the synchronized regime can lead to line tripping and cascading failures, causing large-scale power blackouts \cite{motter_cascade-based_2002,carreras_critical_2002,simonsen_transient_2008}. Apart from physical exterior damage to grid elements, another source of instability are dynamical perturbations caused by short-term input fluctuations of active power (angular instability), a large imbalance of supply and demand of active power (frequency instability), and a shortage of reactive power production and transmission (voltage instability). While the inclusion of renewable energy sources addresses problems associated with conventional power supply such as resource scarcity and climate change, angular stability in transmission grids is jeopardized in particular by the intermittent character of solar and wind power \cite{bevrani_renewable_2010}. For those energy sources, the \emph{temporal} character of input fluctuations is largely responsible for angular instabilities \cite{schmietendorf_stability_2016,anvari_short_2016,bandi_spectrum_2017} and needs to be accounted for in the light of a growing share of such renewables in the energy mix. 

Existing approaches to determine a grid's angular stability are predominantly simulation-based and mainly explore deviations around the synchronized state as a function of specific input perturbations \cite{ji_basin_2014,nagata_node-wise_2014,schmietendorf_self-organized_2014,anvari_short_2016,mitra_multiple-node_2017}. In \cite{manik_network_2017}, power grid susceptibility is defined with respect to a change of the grid's synchronized state through a \emph{permanent} modification of oscillator or transmission line  characteristics. The grid's linear response to these permanent modifications is then analytically predicted. In the following, we present an analytic approach that yields a frequency-resolved measure of an arbitrary grid's susceptibility to \emph{dynamical} input fluctuations. This accounts for the intermittent character of renewable energy sources and extends the notion of grid susceptibility to input fluctuations with arbitrary autocorrelations.

\section{The linear-response ansatz}
Following \cite{filatrella_analysis_2008}, transmission grid dynamics can be written as
\begin{equation}\label{e:kuramoto}
\ddot{\theta}_j(t)=p_j+\xi_j(t)-\alpha \dot{\theta}_j(t)+ \sum_{m=1}^N K_{jm}\sin{\left[  \theta_m(t)-\theta_j(t)\right]}
\end{equation}
for integer $j,m\in[1,N]$, with time dependence explicitly stated. Here $\theta_j(t)$ is the phase of oscillator $j$ at time $t$, and $\alpha$ a normalized fraction of the oscillator's damping torque and moment of inertia, assumed to be uniform across the network. Moreover, we assume a symmetric coupling $K_{jm}=K_{mj}$ between oscillators $j$ and $m$ and a connected graph, as is the case in transmission grids. 
For all practical purposes, the natural frequency $p_j$ of oscillator $j$ is taken to be a positive constant if on average oscillator $j$ injects power into the system (in the case of a generator) and negative if on average it draws power from the grid (i.e., a consumer or a substation) \cite{filatrella_analysis_2008}. Small fluctuations $\xi_j(t)$ around this nominal value signify intermittent power injection or consumption, respectively, with $|\xi_j(t)|\ll |p_j|$ at all times and the time average $\langle\xi_j(t)\rangle_t=0$. Without loss of generality, one can enter a co-rotating reference frame by setting $\sum_{j=1}^N p_j=0$, so that $N^{-1}\sum_{j=1}^{N}\langle\dot{\theta}_j(t)\rangle_t=0$ for the system frequency. Frequency synchronization then occurs if $\theta_j(t)=\theta^*_j$ is constant for all $j\in[1,N]$.

In case of small deviations around the synchronized state, the dynamics of Eq.~\ref{e:kuramoto} can be linearized. For $x_j(t)\equiv \theta_j(t)-\theta^*_j$ and $|x_j(t)|\ll 1$ for all $j\in[1,N]$, one obtains
\begin{equation}\label{e:kuramotolin}
\ddot{x}_j(t)+\alpha \dot{x}_j(t)+\sum_{m=1}^N M_{jm} x_m(t)=\xi_j(t)
\end{equation}
with
\begin{equation}\label{e:Mdef}
M_{jm} =
\begin{cases}
\sum_{n=1}^N K_{jn} \cos{\left(\theta^*_n-\theta^*_j\right)} & \text{if $j=m$}\\
-K_{jm}\cos{\left( \theta^*_m-\theta^*_j \right) } & \text{if $j \neq m$}\, .
\end{cases}
\end{equation}
for the linearized dynamics of the $j$-th oscillator. In the following, let quantities with subscripts be the entries of bold-faced vectors (in the case of one index) or matrices (in the case of two indices) of same designation.  The matrix $\bm{M}$ has real and symmetric entries (Eq.~\ref{e:Mdef}), so that according to the spectral theorem, its right eigenvectors form an orthonormal basis and are the column vectors of an orthogonal matrix $\bm{P}$. Hence $\bm{P}^{-1}=\bm{P}^{T}$, and $\bm{M^d}=\bm{P}^{-1} \bm{M} \bm{P}$ is a diagonal matrix whose diagonal entries are $\bm{M}$'s eigenvalues. 
It follows that with $\bm{x^d}(t)\equiv\bm{P}^{-1} \bm{x}(t)$ and $\bm{\xi^d}(t)\equiv\bm{P}^{-1} \bm{\xi}(t)$, Eq.~\ref{e:kuramotolin} can be written as $\ddot{x}^d_j(t)+\alpha \dot{x}^d_j(t)+ M^d_{jj} x^d_j(t)=\xi^d_j(t)$. Consequently, we obtain a system of $N$ decoupled generalized oscillators. Their linear response in Fourier space to input fluctuations $\bm{\xi^d}(t)$
is $\tilde{x}^d_j(\omega)=\tilde{\chi}^d_{jj}(\omega)\tilde{\xi}^d_j(\omega)$ with the linear response function
\begin{equation}\label{e:linresponsefunc}
\tilde{\chi}^d_{jj}(\omega)=\left[M^d_{jj}-\omega^2-i\alpha \omega\right]^{-1}\, ,
\end{equation}
a complex-valued entry of the diagonal matrix $\bm{\tilde{\chi}^d}(\omega)$. Backtransforming to the original coordinates yields
\begin{equation}\label{e:linresponse}
\tilde{x}_j(\omega)=\textstyle\sum_{m=1}^N\tilde{\chi}_{jm}(\omega)\tilde{\xi}_m(\omega) \, ,
\end{equation}
defining the linear response matrix $\bm{\tilde{\chi}}(\omega)=\bm{P}\bm{\tilde{\chi}^d}(\omega)\bm{P}^{-1}$. 

The matrix $\bm{\tilde{\chi}}(\omega)$ is a frequency-resolved measure of the network's susceptibility to small input perturbations. It is symmetric, as for its transpose $[\bm{\tilde{\chi}}(\omega)]^T=\bm{P}\bm{\tilde{\chi}^d}(\omega)\bm{P}^{-1}=\bm{\tilde{\chi}}(\omega)$ from $\bm{P}^{-1}=\bm{P}^{T}$. Thus the linear response $\tilde{\chi}_{jm}(\omega)\tilde{\xi}_m(\omega)$ of oscillator $j$ to input fluctuations $\tilde{\xi}_m(\omega)$ at oscillator $m$ is the same as the response of $m$ evoked by identical perturbations of $j$. This is the case even if oscillators $m$ and $j$ have different natural frequencies, and a result of the assumed symmetric coupling and uniform damping torques. Note moreover that the entries of $\bm{\tilde{\chi}}(\omega)$ are complex-valued: the real (\emph{reactive}) parts describe in-phase responses, the imaginary (\emph{dissipative}) parts out-of-phase responses. To assess angular stability in the transmission grid, we will concentrate on the absolute values of the responses, as for given perturbations, they quantify the maximum deviations of oscillator phases from their equilibrium values. 
 
\section{The linear response in a regular random graph} 
\begin{figure}[h]
  \centering
    \includegraphics[width=0.9\textwidth]{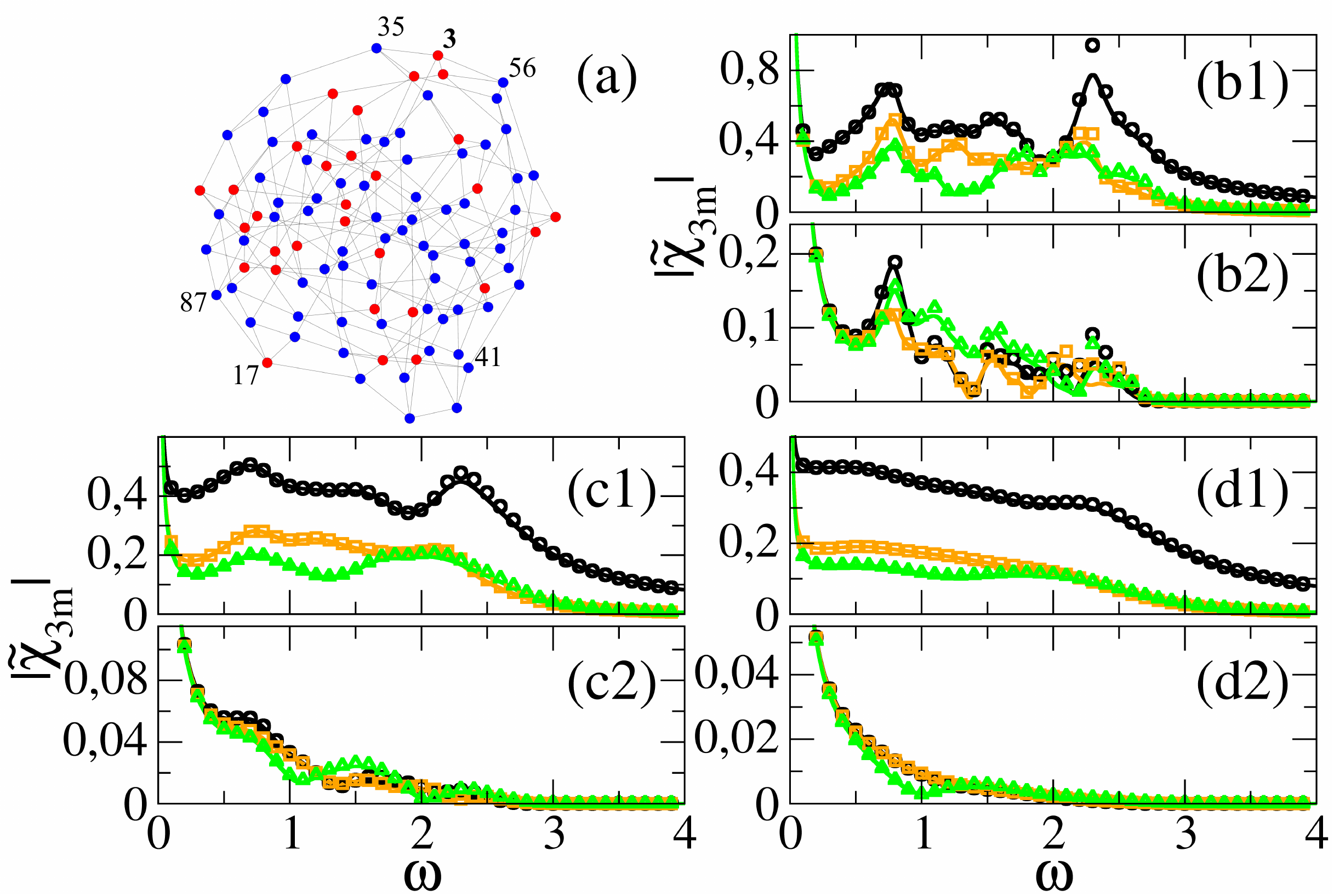}
  \caption{(Color online) Response in regular random graph to input fluctuations at single node. (a) Graph realization with $30$ generators (red), $70$ consumers (blue), degree $k=4$ and $\lambda=1$. Relevant oscillators are indexed. (b1)-(d2) Response of indexed oscillators to input fluctuation $\xi_3(t)=10^{-3}\sin{(\omega t)} $ at generator 3. Solid lines:  predicted absolute value of linear response function in Eq.~\ref{e:linresponse}. Symbols: bisected peak-to-peak amplitudes of respective oscillator phases recorded at $900\leq t \leq 1000$ in numerical integration of Eq.\ref{e:kuramoto}, normalized by the input-fluctuation amplitude. Responses shown for $\alpha=0.2$ [(b1)-(b2)], $\alpha=0.5$ [(c1)-(c2)], $\alpha=1$ [(d1)-(d2)]. (b1), (c1) and (d1): response in generator 3's neighborhood; at generator 3 itself (black lines and circles), at consumer 56 (orange lines and squares) and consumer 35 (green lines and triangles). (b2), (c2) and (d2): response of distant oscillators; at generator 17 (black lines and circles), at consumer 87 (orange lines and squares) and consumer 41 (green lines and triangles). Shortest path length to generator 3 is $5$ in all three cases. Shortest path length from generator 17 to consumer 87 is $1$, and $5$ between consumers 41 and 87.}\label{f:rrg}
\end{figure}

Regular random graphs are randomly connected graphs with the one constraint of each node having the same number of neighbors - or \emph{degree} - $k$. Between complex transmission grids and their simplistic representation as fully connected graphs, we pick regular random graphs as a good middle ground that incorporates structured connections while offering an intuitive setting to explore coupled oscillator dynamics. Moreover, we choose a sparse regular random graph where some oscillators are sufficiently far away from the source of input fluctuations (as quantified by the shortest path length to the source). This allows to investigate which effect input perturbations have on oscillators at a varying distance. Lastly, we set in Eq.~\ref{e:kuramoto} $K_{jm}=\lambda$ if oscillators $j$ and $m$ are connected, and $K_{jm}=0$ otherwise, yielding uniform coupling strengths. For the particular graph realization in Fig.~\ref{f:rrg}(a), the critical threshold for frequency synchronization is $\lambda^*\approx 0.50$.  We therefore choose $\lambda=1$ in numerical integration of Eq.~\ref{e:kuramoto} to obtain the graph's steady state and investigate its response to small perturbations at a single generator. All numerical integration is performed with the Heun scheme with step size $10^{-2}$ and statistics recorded between $900\leq t\leq 1000$.

Figures~\ref{f:rrg}(b1)-(d2) show, for fluctuating input at generator 3, typical plots of the response of selected oscillators in frequency space. In all cases, our ansatz predicts very well the oscillators' responses. As expected, the response is the weaker the larger is the damping torque $\alpha$ and the farther is the considered oscillator from the generator with fluctuating input. For all considered oscillators and sufficiently small $\alpha$, we find several local extrema at non-zero frequencies, in contrast to the unique maximum of the absolute value of the linear response of an isolated perturbed oscillator. The reason is that the matrix entries $\tilde{\chi}_{jm}(\omega)=\sum_{l=1}^N P_{jl}P_{ml}\tilde{\chi}^d_{ll}(\omega)$ are a linear combination of the linear response functions of uncoupled oscillators in Eq.~\ref{e:linresponsefunc}. This yields the observed non-trivial behavior, in particular this of the perturbed generator 3 itself [black data in Figs.~\ref{f:rrg}(b1), (c1) and (d1)].

In the vicinity of generator 3 with fluctuating input, the responses of two selected consumer neighbors 35 and 56 differ from that of 3 as well as from each other, even though the two consumers possess the same number of generator and consumer neighbors [Figs.~\ref{f:rrg}(b1), (c1) and (d1)]. For oscillators 17, 41 and 87 which are more distant from the source of input fluctuations, we similarly find that the responses of consumers 41 and 87 (both distant from generator 3 and from each other) differ significantly, even though - as for consumers 35 and 56 - they possess the same neighborhood composition and natural frequency. This indicates that mean-field approaches like the active-neighborhood ansatz in \cite{wieland_mean-field_2017} are insufficient to capture how input perturbations around the equilibrium propagate through the oscillator network. Further evidence is given by the observation that the response of a directly linked generator-consumer pair is very similar even in a remote graph region [orange and black data in Figs.~\ref{f:rrg}(b2), (c2) and (d2)], despite the difference in natural frequencies and neighborhood composition. Hence, the number of short path lengths between two oscillators (and the ensuing notion of "proximity" in the graph) seems to determine whether they respond similarly to input fluctuations. 

Figures~\ref{f:rrg}(b1)-(d2) also suggest that particularly for sufficiently low damping torques, the peaks in the response to input fluctuations can differ from oscillator to oscillator. Hence for input fluctuations at generator 3, the angular stability in this particular power grid architecture could be compromised at several frequency components. The fluctuating input at another generator propagates through different graph sections to the previously considered oscillators, and will thus evoke different responses from them. Consequently, the response of an oscillator depends on the location of the source of input fluctuations. Furthermore, Eq.~\ref{e:linresponse} predicts equally well responses to in-phase input fluctuations at multiple, arbitrarily chosen generators through (not shown).
%
%
\section{Summary and conclusions}
Here we present, for arbitrary grid architectures, an analytical framework to determine transmission grid susceptibilities to small input fluctuations. To this end, we make use of the linear response theory for generalized oscillators and find non-trivial response functions of selected grid oscillators, in particular for sufficiently small damping torques of the rotating machines that model power generators and substations. For a given grid topology, one can therefore analytically predict frequency components in input fluctuations that compromise angular stability in the grid. This is of particular relevance with respect to the inclusion of renewable energies with a strong intermittent character, i.e, wind and solar power. Identified harmful frequency components could be potentially addressed by the inclusion of appropriate storage elements. The proposed approach generalizes to non-uniform symmetric coupling strengths as well as responses to fluctuating power consumptions. It assumes uniform damping torques and in-phase input perturbations to give a coarse-grained analytical understanding of the susceptibility of a given grid architecture to input fluctuations. For future work, a generalization to more heterogeneous oscillator parameters and input characteristics is in order to account for more realistic grid setups. 

\begin{acknowledgments}
The authors would like to thank F. Daviaud, F. Suard, M. Velay and M. Vinyals for useful discussions. Funding by CEA under the NTE NESTOR grant is gratefully acknowledged. Additionally, S.W. acknowledges funding of project 330 in the Enhanced Eurotalents program.
\end{acknowledgments}

\newpage
\renewcommand\refname{\textbf{References}}
\bibliographystyle{unsrt}
\bibliography{draft}

\begin{thebibliography}{10}

\bibitem{kuramoto_chemical_1984}
Yoshiki Kuramoto.
\newblock {\em Chemical {Oscillations}, {Waves}, and {Turbulence}}, volume~19
  of {\em Springer {Series} in {Synergetics}}.
\newblock Springer Berlin Heidelberg, Berlin, Heidelberg, 1984.
\newblock DOI: 10.1007/978-3-642-69689-3.

\bibitem{filatrella_analysis_2008}
G.~Filatrella, A.~H. Nielsen, and N.~F. Pedersen.
\newblock Analysis of a power grid using a {Kuramoto}-like model.
\newblock {\em The European Physical Journal B}, 61(4):485--491, February 2008.

\bibitem{nishikawa_comparative_2015}
Takashi Nishikawa and Adilson~E Motter.
\newblock Comparative analysis of existing models for power-grid
  synchronization.
\newblock {\em New Journal of Physics}, 17(1):015012, January 2015.

\bibitem{acebron_kuramoto_2005}
Juan~A. Acebrón, Luis~L. Bonilla, Conrad J.~Pérez Vicente, Félix Ritort, and
  Renato Spigler.
\newblock The {Kuramoto} model: {A} simple paradigm for synchronization
  phenomena.
\newblock {\em Reviews of modern physics}, 77(1):137, 2005.

\bibitem{arenas_synchronization_2008}
Alex Arenas, Albert Díaz-Guilera, Jurgen Kurths, Yamir Moreno, and Changsong
  Zhou.
\newblock Synchronization in complex networks.
\newblock {\em Physics Reports}, 469(3):93--153, December 2008.

\bibitem{dorfler_synchronization_2014}
Florian Dörfler and Francesco Bullo.
\newblock Synchronization in complex networks of phase oscillators: {A} survey.
\newblock {\em Automatica}, 50(6):1539--1564, June 2014.

\bibitem{gupta_kuramoto_2014}
Shamik Gupta, Alessandro Campa, and Stefano Ruffo.
\newblock Kuramoto model of synchronization: equilibrium and nonequilibrium
  aspects.
\newblock {\em Journal of Statistical Mechanics: Theory and Experiment},
  2014(8):R08001, 2014.

\bibitem{rodrigues_kuramoto_2016}
Francisco~A. Rodrigues, Thomas K.~DM. Peron, Peng Ji, and Jürgen Kurths.
\newblock The {Kuramoto} model in complex networks.
\newblock {\em Physics Reports}, 610:1--98, January 2016.

\bibitem{motter_cascade-based_2002}
Adilson~E. Motter and Ying-Cheng Lai.
\newblock Cascade-based attacks on complex networks.
\newblock {\em Physical Review E}, 66(6), December 2002.

\bibitem{carreras_critical_2002}
B.~A. Carreras, V.~E. Lynch, I.~Dobson, and D.~E. Newman.
\newblock Critical points and transitions in an electric power transmission
  model for cascading failure blackouts.
\newblock {\em Chaos: An Interdisciplinary Journal of Nonlinear Science},
  12(4):985--994, December 2002.

\bibitem{simonsen_transient_2008}
Ingve Simonsen, Lubos Buzna, Karsten Peters, Stefan Bornholdt, and Dirk
  Helbing.
\newblock Transient {Dynamics} {Increasing} {Network} {Vulnerability} to
  {Cascading} {Failures}.
\newblock {\em Physical Review Letters}, 100(21), May 2008.

\bibitem{bevrani_renewable_2010}
H.~Bevrani, A.~Ghosh, and G.~Ledwich.
\newblock Renewable energy sources and frequency regulation: survey and new
  perspectives.
\newblock {\em IET Renewable Power Generation}, 4(5):438, 2010.

\bibitem{schmietendorf_stability_2016}
Katrin Schmietendorf, Joachim Peinke, and Oliver Kamps.
\newblock On the stability and quality of power grids subjected to intermittent
  feed-in.
\newblock {\em arXiv preprint arXiv:1611.08235}, 2016.

\bibitem{anvari_short_2016}
M~Anvari, G~Lohmann, M~Wächter, P~Milan, E~Lorenz, D~Heinemann, M~Reza~Rahimi
  Tabar, and Joachim Peinke.
\newblock Short term fluctuations of wind and solar power systems.
\newblock {\em New Journal of Physics}, 18(6):063027, June 2016.

\bibitem{bandi_spectrum_2017}
M.~M. Bandi.
\newblock Spectrum of {Wind} {Power} {Fluctuations}.
\newblock {\em Physical Review Letters}, 118(2), January 2017.

\bibitem{ji_basin_2014}
Peng Ji and Jürgen Kurths.
\newblock Basin stability of the {Kuramoto}-like model in small networks.
\newblock {\em The European Physical Journal Special Topics},
  223(12):2483--2491, October 2014.

\bibitem{nagata_node-wise_2014}
Motoki Nagata, Naoya Fujiwara, Gouhei Tanaka, Hideyuki Suzuki, Eiichi Kohda,
  and Kazuyuki Aihara.
\newblock Node-wise robustness against fluctuations of power consumption in
  power grids.
\newblock {\em The European Physical Journal Special Topics},
  223(12):2549--2559, October 2014.

\bibitem{schmietendorf_self-organized_2014}
Katrin Schmietendorf, Joachim Peinke, Rudolf Friedrich, and Oliver Kamps.
\newblock Self-organized synchronization and voltage stability in networks of
  synchronous machines.
\newblock {\em The European Physical Journal Special Topics},
  223(12):2577--2592, October 2014.

\bibitem{mitra_multiple-node_2017}
Chiranjit Mitra, Anshul Choudhary, Sudeshna Sinha, Jürgen Kurths, and Reik~V.
  Donner.
\newblock Multiple-node basin stability in complex dynamical networks.
\newblock {\em Physical Review E}, 95(3), March 2017.

\bibitem{manik_network_2017}
Debsankha Manik, Martin Rohden, Henrik Ronellenfitsch, Xiaozhu Zhang, Sarah
  Hallerberg, Dirk Witthaut, and Marc Timme.
\newblock Network susceptibilities: {Theory} and applications.
\newblock {\em Physical Review E}, 95(1), January 2017.

\bibitem{wieland_mean-field_2017}
Stefan Wieland, Simone~Blanco Malerba, Sébastien Aumaitre, and Hervé
  Bercegol.
\newblock Mean-field approach for frequency synchronization in complex networks
  of two oscillator types.
\newblock {\em arXiv preprint arXiv:1712.04352}, 2017.

\end{thebibliography}
\end{document}